\documentclass[11pt]{article}
\usepackage{verbatim,cancel}

\title{Still Simpler Way of Introducing Interior-Point method for Linear
Programming}

\author{Sanjeev Saxena\thanks{E-mail: ssax@iitk.ac.in}\\
Dept. of Computer Science and
Engineering,\\ Indian Institute of Technology,\\
Kanpur, INDIA-208 016}

\date{\today}

\begin{document}
\maketitle

\subsection*{\centering{Abstract}}

Linear Programming is now included in Algorithm undergraduate and
postgraduate courses for Computer Science majors. It is possible to teach
interior-point methods directly with just minimal knowledge of Algebra
and Matrices. 

\section{Introduction}

Terlaky[3] and Lesaja[1] have suggested simple ways to teach
interior-point methods. In this paper a still simpler way is being
suggested. Most material required to teach interior-point methods is
available in popular text books[2,4]. However, these books assume
knowledge of Calculus, which is not really required. In this paper, it is
suggested if appropriate material is selected from these books then it
becomes very easy to teach interior-point methods as the first or only
method for Linear programming in Computer Science Courses.

Canonical Linear Programming Problem is to\\ minimise $cx^T$ subject to
$Ax=b$ and $x\ge 0$.\\

Here $A$ is an $n* m$ matrix, $b$ and $c$ are $n$-dimensional and
$x$ is an $m$-dimensional vector. 

Remark~1. maximise $cx^T$ is equivalent to minimise $-cx^T$.

Remark~2. Constraints of type $\alpha_1x_1+... +\alpha_nx_n\le \beta$ can
be replaced by $\alpha_1x_1+... +\alpha_nx_n+\gamma = \beta$ with a new
(slack) variable $\gamma\ge 0$. Similarly constraints of type
$\alpha_1x_1+... +\alpha_nx_n\ge \beta$ can be replaced by
$\alpha_1x_1+... +\alpha_nx_n-\gamma = \beta$ with (surplus) variable
$\gamma\ge 0$.

Thus, we assume that there are $n$ constraints and $m$ variables, with
$m>n$ (more variables and fewer constraints)--- basically slack or
surplus are added or subtracted to convert inequalities into equalities.

We first use pivoting to make first term of all but the first equation as
zero. Basically, we multiply $i$th equation by $-{a_{11}}/{a_{i1}}$ and
subtract the first equation. In similar way we make first two terms of
all but the first two equations as zero-- multiply $i$th equation (for
$i\neq 2$) by $-{a_{22}}/{a_{21}}$ and subtract the second equation. And
so on. In case, if in any equation all coefficients become zero, we drop
those equations. As a result, in the end all remaining equations will be
linearly independent. Or the resulting matrix will have full row rank.

Remark. We may have to interchange two columns (interchange two
variables), in case, for example, if a diagonal term of an equation
becomes zero.

From convexity, it is sufficient to obtain a locally optimal solution, as
local optimality will imply global optimality.

We consider another problem, the ``dual problem'' which is \\
maximise $by^T$ subject to $A^Ty+s=c$, with 
slack variables $s\ge 0$ and variables $y$ are unconstrained.\\
{\textbf{Claim~1}} 
$by^T\le cx^T$. The equality will hold if and only if, $s_ix_i=0$ for
all $i$s.

Remark. Thus if value of both primal and dual are the same, then both are
optimal.

Proof. $s=c-A^Ty$, or $x^Ts=x^Tc- x^T(A^Ty)=c^Tx-(x^TA^T)y=c^Tx-b^Ty$. 
As $x,s\ge 0$, we have $x^Ts\ge 0$ or $c^Tx\ge b^Ty$.

Equality will hold if $x^Ts=0$ or $\sum_i s_ix_i=0$ but as $s_i,x_i\ge
0$, we want each term (product) $s_ix_i=0$. $[]$

Thus, if we are able to find a solution of following equations (last one
is not linear, else, an inversion of matrix would have been sufficient),
we will be getting optimal solutions of both the original and the dual
problems.\\
$ Ax=b, A^Ty+s=c, x_is_i=0$\\
subject to $x\ge 0, s\ge 0$.

We will relax the last condition to get something like (duality gap):\\
$x_is_i\approx \mu$\\ with parameter $\mu\ge 0$. Thus, we will be solving
(the exact last equation will be derived in the next section):\\
$ Ax=b, A^Ty+s=c, x_is_i\approx \mu$ subject to $x\ge 0, s\ge 0$.

Remark. Thus, $by^T-cx^T \approx m\mu$. If $\mu$ is very small, then in
the case of rationals, the solution will be exact.

\section{Use of Newton Raphson Method}

We will use the Newton-Raphson method[2]. Let us choose the next
values as $x+h,y+k,s+f$. Then we want:\\
(1) $A(x+h)=b$ or $Ax+Ah=b$
but as $Ax=b$, we get 
$Ah=0$.\\
(2) $A^T(y+k)+(s+f)=c$, from $A^Ty+s=c$, we get 
$A^Tk+f=c-A^Ty-s=0$ or $A^Tk+f=0$\\
(3) $(x_i+h_i)(s_i+f_i)\approx \mu$ or
$x_is_i+h_is_i+f_ix_i+h_if_i\approx \mu$.
Or approximately, $x_is_i+h_is_i+f_ix_i=\mu$ (neglecting the non-linear
$h_if_i$ term). Thus, the equation we will be solving is\\
$h_is_i+f_ix_i=\mu -x_is_i$

Thus, we have a system of linear equations for $h_i,k_i,f_i$. We next
show that these can be solved by ``inverting'' a matrix.

But first observe that from the third equation,
\\{\textbf{Observation~1}} 
$(x_i+h_i)(s_i+f_i)=\mu+h_if_i$ $[]$\\
{\textbf{Theorem~1}}
Following equations have a unique solution:\\
(1) $Ah=0$\\
(2) $A^Tk+f=0$\\
(3) $h_i s_i+f_i x_i=\mu-x_is_i$

Proof. We will follow Vanderbei[4] and use capital letters (e.g. $X$) in
this proof (only) to denote a diagonal matrix with entries of the
corresponding row vector (e.g. in $X$ the diagonal entries will be
$x_1,x_2,... ,x_m$). We will also use $e$ to denote a column vector of
all ones (usually of length $m$).

Then in the new notation, the last equation is:\\
$S h+Xf=\mu e-XSe$
Let us look at this equation in more detail.\\

$Sh+Xf = \mu e-XSe$

or

$h+ S^{-1}Xf = S^{-1}\mu e- S^{-1}XSe$ (pre-multiply by $S^{-1}$) 

or

$h+S^{-1}Xf = \mu S^{-1}e -X {S^{-1}}{S} e$   (diagonal
matrices commute)

or

$h+S^{-1}Xf = \mu S^{-1}e -x$ (as  $Xe=x$)

or

$Ah+ AS^{-1}Xf = \mu AS^{-1}e-Ax$  (pre-multiply by $A$) 

or

$AS^{-1}Xf = \mu AS^{-1}e-b$ (but $Ax=b$ and $Ah=0$) 

or

$-AS^{-1}XA^Tk = \mu AS^{-1}e-b$ (using  $f=-A^Tk$)

or

$b-AS^{-1}e = (AS^{-1}XA^T)k$

As $XS^{-1}$ is diagonal with positive items and as $A$ has full rank, thus
$AS^{-1}XA^T$ is invertible (see appendix). The last
equation can thus be used to get the value of matrix $k$ after inverting
the matrix $AS^{-1}XA^T$, or\\ $k=(AS^{-1}XA^T)^{-1}(b-AS^{-1}e)$

Then we can find $f$ from $f=-A^Tk$.

And to get $h$ we use the equation:
$h+S^{-1}Xf = \mu S^{-1}e -x$, i.e.,\\
$h= -XS^{-1}f+\mu S^{-1}e-x$

Thus, the above system has a unique solution. $[]$\\
\textbf{Claim 2}
$\sum_i h_i f_i=0$ or equivalently $h^Tf=f^Th=0$

Proof As $A^Tk+f=0$, we get $h^TA^Tk+h^Tf=0$ but $h^TA^T=(Ah)^T=0$, hence
$h^Tf=0$ follows. $[]$

\section{Invariants in each Iteration}

We will maintain following invariants:\\
(1) $Ax^T=b$, with $x>0$ (strict inequality)\\
(2) $A^Ty+s=c$ with $s>0$ (strict inequality)\\
(3) If $\mu$
is the ``approximate duality gap'' then $\sigma\le 2/3< \sqrt{3}-1$
where $\sigma^2=\sum_i (({x_i s_i}/{\mu})-1)^2$.

At end of this iteration we want duality gap $\mu'\le (1-\delta)\mu$. We
will see that $\delta$ can be chosen as $\delta=\Theta(1/\sqrt{m})$.

We first show that strict inequality invariants hold (in $\sigma'$ we
have $x+h,s+f$ and same $\mu$):\\
\textbf{Fact~1} 
If $\sigma'<1$ then $x+h>0$ and $s+f>0$

Proof. We first show that the product $(x_i+h_i)(s_i+f_i)$ is term-wise
positive. From Observation~1, $(x_i+h_i)(s_i+f_i)=\mu+h_is_i$.

From $\sigma'<1$ we get $\sigma'^2< 1$. But (using Observation~1):\\
$\sigma'^2=\sum_i ({(x_i+h_i)( s_i+f_i)}/{\mu}-1)^2=\sum_i
({h_is_i}/{\mu})^2<1$\\ As the sum is at most one, it follows that each
term of the summation must be less than one, or $| {h_if_i}/{\mu}|
<1$ or $-\mu< h_if_i<\mu$. In particular $\mu+h_if_i>0$.

Thus the product $(x_i+h)(s_i+f)$ is term-wise positive.

Assume for contradiction that both $x_i+h_i<0$ and $s_i+f_i<0$. But as
$s_i>0$ and $h_i>0$, we have $s_i(x_i+h_i)+x_i(s_i+f_i)<0$, or
$\mu+x_is_i<0$. Which is impossible as $\mu,x_i,s_i$ are all
non-negative, a contradiction. $[]$

We have to still show that the ``approximate duality gap'' $\mu$
decreases as desired.

Let us define three new variables:\\
$H_i = h_i\sqrt{{s_i}/{x_i\mu}}$ and \\
$F_i = f_i\sqrt{{x_i}/{s_i\mu}}$

Observe that $\sum_i H_iF_i=\sum {h_if_i}/{\mu}=0$ (see Claim~2).

$H_i+F_i = h_i\sqrt{{s_i}/{x_i\mu}}+ f_i\sqrt{{x_i}/{s_i\mu}}$\\
$= \sqrt{{1}/{x_is_i\mu}} (h_is_i+f_ix_i)$\\
$=  \sqrt{{1}/{x_is_i\mu}} (\mu-x_is_i)$\\
$=  \sqrt{{\mu}/{x_is_i}} (1- {x_is_i}/{\mu})$\\
$= -\sqrt{{\mu}/{x_is_i}} (-1+{x_is_i}/{\mu})$

From, the proof of Fact~1 we also observe that $\sigma'^2=\sum_i
({h_if_i}/{\mu})^2$, or $\sigma'^2=\sum_i (H_iF_i)^2$

And finally\\

$\sigma'^2 = \sum_i (H_iF_i)^2$\\ 
$\le \sum_i {(H_i^2+F_i^2)^2}/{4}$  (using AM-GM
inequality)\\
$\le {1}/{4}(\sum_i (H_i^2+F_i^2))^2$ (more positive
terms)\\
$\le {1}/{4}(\sum_i (H_i+F_i)^2)^2$ (from 
Claim~2)\\
$= {1}/{4}(\sum_i {\mu}/{x_is_i}
({x_is_i}/{\mu}-1)^2)^2$ \\
$\le (\max {\mu}/{x_is_i})^2{1}/{4}(\sum
({x_is_i}/{\mu}-1)^2)^2$\\
$\le {\sigma^4}/{4}(\max {\mu}/{x_is_i})^2$

As $\sigma^2=\sum ({x_is_i}/{\mu}-1)^2$, each individual
term is at most $\sigma^2$ or\\
$|{x_is_i}/{\mu}-1|\le \sigma$\\
Thus, 
$-\sigma \le {x_is_i}/{\mu}-1 \le \sigma$ or
$1-\sigma \le {x_is_i}/{\mu} \le 1+\sigma$

In particular ${\mu}/{x_is_i}\le 1/{(1-\sigma)}$ or \\$\max
{\mu}/{x_is_i}\le 1/{(1-\sigma)}$\\ 

Thus, $\sigma'^2\le (1/{(1-\sigma)})^2{\sigma^4}/{4}$ or $\sigma'\le (1/2)
{\sigma^2}/{(1-\sigma)}$

We summarise our observations as:\\
\textbf{Observation~2}
$\sigma'\le ({1}/{2}) {\sigma^2}/({1-\sigma})$

For $\sigma'<1$, ${\sigma^2}/({1-\sigma})<2$ or $\sigma^2<2-2\sigma$ or
$\sigma^2+2\sigma-2<0$ or $\sigma=\sqrt{3}-1$. 

Remark. Thus $\sigma\le 2/3$ is more than enough.

Let us finally try to get bounds on $\delta$ (and hence $\mu$).
Let us assume $\mu'=\mu(1-\delta)$ then if $\sigma''$ corresponds to
$x+h,s+f$ and $\mu'$ we have\\
$\sigma''^2 =
\sum_i({(x_i+h_i)(s_i+f_i)}/{\mu(1-\delta)}-1)^2$\\
$=
\sum_i({(x_i+h_i)(s_i+f_i)-\mu(1-\delta)}/{\mu(1-\delta)})^2$\\
$=
\sum_i({(x_i+h_i)(s_i+f_i)-\mu}/{\mu(1-\delta)}+{\delta}/{1-\delta})^2$\\
$=
\sum_i({h_if_i}/{\mu(1-\delta)}+{\delta}/{1-\delta})^2$
 (From Observation~1)\\
$= {1}/{(1-\delta)^2}
\sum_i({h_if_i}/{\mu}+\delta )^2$\\
$= {1}/{(1-\delta)^2}(
\sum_i({h_if_i}/{\mu})^2+m\delta^2+{2\delta}/{\mu}\sum
h_if_i )$\\
$= {1}/{(1-\delta)^2}(
\sum_i({h_if_i}/{\mu})^2+m\delta^2 )$ { (From Claim~2)}\\
$= {1}/{(1-\delta)^2}(\sigma'^2+m\delta^2 )$\\

Thus observe that\\
\textbf{Observation~3}
$\sigma''= ({1}/{1-\delta})\sqrt{\sigma'^2+m\delta^2}$\\

We want to choose $\delta$ such that $\sigma''\le 2/3$. As $\sigma''=
({1}/{1-\delta})\sqrt{\sigma'^2+m\delta^2}\le
({1}/{1-\delta})\sqrt{m\delta^2}={\delta \sqrt{m}}/({1-\delta})$

We want ${\delta \sqrt{m}}/{1-\delta}<2/3$ or
${\delta}/({1-\delta})<{2}/{3\sqrt{m}}$. We can thus choose
$\delta={1}/{4\sqrt{m}}$. 

\section*{Summary}

Let us assume that initial duality gap is $\mu_0$ and final duality gap
is $\mu_f$, as after each iteration, $\mu'\le (1-\delta)\mu$, thus after
$r$ iterations, $\mu_f \le (1-\delta)^r \mu_0$, or \\$\log
{\mu_0}/{\mu_f}=-r \log (1-\delta)\approx -r (-\delta)$\\

or \\$r=O({1}/{\delta}\log {\mu_0}/{\mu_f})=
O(\sqrt{m}\log {\mu_0}/{\mu_f})$\\

As $1-\sigma\le {x_is_i}/{\mu}\le 1+\sigma$, we have (in last
inequality we use $\sigma<2/3$).
\\$\mu (1-\sigma)\le x_is_i \le \mu(1+\sigma)< {5}/{3}\mu$\\

Thus, when $\mu$ becomes very small, even the products $x_is_i$s will be
very small.
The above method will give a polynomial time algorithm even if
$\mu_0=2^{m^{O(1)}}$ and $\mu_f={1}/{2^{m^{O(1)}}}$.

To find an initial solution, we can use the method suggested 
by Bertsimas and Tsitsiklis[5,p430], which is described, for
completeness in Section 5.

\section{Initial Solution}

This section is based on description of Mehlhorn[6].

Let us first assume that there is a number (say) $W$ such that there is
an optimal solution $x^*$ for which each $|x_i| \leq W$; we will see
later (see Section 6) how to find such a number in case all enteries of
$A$ and $b$ are integers. If $e$ is a column vector (of length $m$) of
all ones, then $e^Tx^*< mW$.

Thus [5,p430] (see also [7,p128-129]) an optimal solution of the
problem\\ minimise $cx^T$ subject to $Ax=b$, $e^Tx<mW$ and $x\geq 0$.\\
will also be a solution of the original problem (without $e^Tx<mW$
constraint). Let us replace (scale) variables $x_i$ by $(mWx'_i)/(m+2)$
then the problem becomes:\\
minimise $\frac{mW}{m+2}cx'^T$ subject to $Ax'=d$, $e^Tx'<m+2$
and $x'\geq 0$ with $d=b(m+2)/(mW)$ 

We add a new variable $x'_{m+1}$ and replace $e^Tx'<m+2$ by
$e^Tx'+x'_{m+1}=m+2$ (with $x_{m+1}\geq 0$). Or dropping primes, the
problem is equivalent to:\\
minimise $cx^T$ subject to $Ax=d$, $e^Tx+x_{m+1}=m+2$
and $x\geq 0$.

Consider a starting solution $x_0$ s.t. all components of $x_0$ are
strictly positive (say all $x_i=1$ or $x=e$, $x_{m+1}=1$). Define a
vector $\rho=d-Ae$. Let $x_{m+2}$ be one more new variable. Then
$Ax+x_{m+2}\rho=d$ with $x\geq 0, x_{m+2}\geq 0$ has a solution with
$x=e$ and $x_{m+1}=x_{m+2}=1$. For this choice,
$e^Tx+x_{m+1}+x_{m+2}=m+2$ is also true. We want a solution in which
$x_{m+2}=0$. Thus, we try to minimise $cx^T+Mx_{m+2}$ for a large $M$.

Remark: It is sufficient to choose $M> mW* max |c_i|$.

We thus consider the artifical primal problem:\\
minimise $cx^T+Mx_{m+2}$\\
subject to\\
$Ax+\rho x_{m+2}=d$,\\
$e^Tx+x_{m+1}+x_{m+2}=m+2$\\
and $x\geq 0, x_{m+1}\geq 0$ and $x_{m+2}\geq 0$.

Remark: If in optimal solution $x_{m+2}>0$, then either there is no
feasible solution, or the value of $M$ chosen was not large enough.

The dual problem (with new dual variable $y_{n+1}, s_{m+1}$ and
$s_{m+2}$) is:\\
maximise $dy^T+(m+2)y_{n+1}$ subject to\\ $A^Ty+e y_{n+1}+s=c$,\\
$\rho^T y+y_{n+1}+s_{m+2}=M$\\ $y_{n+1}+s_{m+1}=0$ with slack
variables $s\geq 0, s_{m+1}>0, s_{m+2}>0$ and variables $y$ are
unconstrained.

To get an initial solution, as $x_{m+1}=1$, we try
$s_{m+1}=\mu/x_{m+1}=\mu$. Then from the last equation
$y_{n+1}=-s_{m+1}=-\mu$. The simplest choice will be to choose all other
$y=0$ then from first equation $s=c+e\mu$ which is again a positive
number (if $\mu$ is larger than all $-c_i$s). To satisfy the second
equation we must choose $s_{m+2}=M-y_{n+1}=M+\mu$. Observe that all slack
variables are positive (provided $\mu$ is large enough).

For this choice, $x_is_i=c_i+\mu$ or $(x_is_i/\mu)-1=c_i/\mu$;
$x_{m+1}s_{m+1}=\mu$ and $(x_{m+2}s_{m+2}/\mu)-1=M/\mu$. Thus
$\sigma^2=(1/\mu^2)(\sum c^2_i+M^2)$. We can make $\sigma^2<1/4$ by
choosing $\mu^2\geq 4\sum (c^2_i+M^2)$.

\section{Integer Case}

This section assumes some more knowledge of algebra-- determinants and
Cramer's rule and some knowledge of geometry. 

If $A$ is an $n* n$ matrix then det$|A|=\sum_\pi
(-1)^{\pi}a_{1\pi(1)}a_{2\pi(2)}... a_{n\pi(n)}$ will be sum of all
possible (products) of permutations $\pi$ (with appropriate sign).
Clearly\\ det$|A|=\sum_\pi (-1)^{\pi}a_{1\pi(1)}a_{2\pi(2)}...
a_{n\pi(n)}\leq \sum_\pi \left|a_{1\pi(1)}a_{2\pi(2)}...
a_{n\pi(n)}\right|$.
If\footnote{ (see e.g. [5,pp 373-374], [7,p75] or [8,pp
43-44])} each $a_{ij}\leq U$, then det$|A|\leq n!U^n$.

Cramer's rule says that solution of equation $Ax=b$ (for $n* n$
non-singular matrix $A$) is $x_i=det(A_i)/det(A)$ where $A_i$ is
obtained by replacing $i$th column of $A$ by $b$. 

As all constraints are linear, solution space will be a convex polytope
and (by convexity) for optimal solution it is sufficient to look at
corner points. At each corner point exactly $n$ components of $x$ will be
non-zero; remaining $m-n$, $x_i$ will be zero. Thus, at optimal solution
$x_i=det(A'_i)/det(A')$, where $A'$ is obtained by keeping only (some)
$n$ columns of $A$. If we also assume that each $|b_i|\le U$, the maximum
value of the determinant can be $n!U^n$.  If all enteries are integers,
then determinant has to be at least one if it is non-zero.

Thus, each $x_i$ is between $1/(n!U^n)$ and $n!U^n$. Or we can choose
$W=n!U^n<(nU)^n$.

\subsection*{Acknowledgement}

This work was inspired by an informal lecture given by Nisheeth Vishnoi
at IIT Kanpur. 
I also wish to thank students of CS602 (2014-2015 batch)
for their helpful comments and questions when I was teaching this
material. Thanks also to Kurt Mehlhorn for pointing out that method for
finding initial solution (of an earlier version) may not work and for his suggestion of using
method of [5,p430] instead.

\bibliographystyle{acm}
\bibliography{general}

\subsection*{Appendix: Result from Algebra} 

Assume that $A$ is $n*  m$ matrix and rank of $A$ is $n$, with $n<m$.
Then all $n$ rows of $A$ are linearly independent. Or
$\alpha_1A_1+\alpha_2A_2+{...} +\alpha_nA_n=0$ (here $0$ is a row
vector of size $m$) has only one solution $\alpha_i=0$. Thus,
if $x$ is any $1*  n$ matrix (a column vector of size $n$), then
$xA=0$ implies $x=0$.

As $A$ is $n*  m$ matrix, $A^T$ will be $m*  n$ matrix. The
product $AA^T$ will be an $n*  n$ square matrix. Let $y^T$ be an
$n*  1$ matrix (or $y$ is a row-vector of size $n$). 

Consider the equation $(AA^T)y^T=0$. Pre-multiplying by $y$ we get
$yAA^Ty^T=0$ or $(yA)(yA)^T=0$ or the dot product $<yA,yA>=0$ which, for
real vectors (matrices) means, that each term of $yA$ is (individually)
zero, or $y$ is identically zero.

Thus, the matrix $AA^T$ has rank $n$ and is invertible.

Also observe that if $X$ is a diagonal matrix (with all diagonal entries
non-zero) and if $A$ has full row-rank, then $AX$ will also have full
row-rank. Basically if entries of $X$ are $x_1,x_2,{...} ,x_n$ then
the matrix $AX$ will have rows as $x_1A_1,x_2A_2,{...} ,x_nA_n$ (i.e.,
$i$th row of $A$ gets scaled by $x_i$). If rows of $AX$ are not
independent then there are $\beta$s (not all zero) such that:\\
$\beta_1x_1A_1+\beta_2x_2A_2+{...} +\beta_nx_nA_n=0$, or there are
$\alpha$s (not all zero) such that:\\
$\alpha_1A_1+\alpha_2A_2+{...} +\alpha_nA_n=0$ with $\alpha_i=\beta_ix_i$.

\end{document}